# DISPERSIVE THEORY OF CHARMONIUM ON THE LATTICE

A. BOCHKAREV [a]

*Theoretical Physics Institute, University of Minnesota, Minneapolis, MN 55455*

Physical contribution of the high-energy part of hadronic spectrum is incorporated to describe the short-distance part of the correlator of heavy-quark currents obtained by quenched Monte-Carlo on a $8^3 * 16$- and $16^3 * 32$-lattices for $\beta = \{6, 6.3\}$. The lattice artifacts in the short-distance behavior of that correlator are isolated. The physical short-distance part of the correlator is fitted by the relevant expressions of perturbative QCD, which allows one to obtain the renormalized charmed quark mass with rather high accuracy $m_c^{\overline{MS}}(m_c) = 1.22(5)\, GeV$.

## 1 Correlator of the heavy-quark currents

The correlator of the vector currents of charmed quarks:

$$\Pi(q^2)_{\mu\nu} = i \int dx e^{iqx} < 0 |T\{j_\mu(x) j_\nu(0)\}| 0 > \qquad (1)$$

where $j_\mu = \bar{c}\gamma_\mu c$ can be used to calculate the parameters of perturbative QCD, such as the renormalized charmed-quark mass $m_c$ and the strong coupling constant $\alpha_s(4m_c^2)$ [1]. On one hand this correlator satisfies the standard dispersion relation:

$$\Pi(q^2)_{\mu\nu} = (q_\mu q_\nu - g_{\mu\nu} q^2) q^2 \int ds\, \frac{\rho(s)}{s^2(s+q^2)} + d_1 g_{\mu\nu} + d_2 (q_\mu q_\nu - g_{\mu\nu} q^2) \qquad (2)$$

with the singular subtraction constants $d_{1,2}$. The spectral density $\rho(s)$ is proportional to the observable inclusive cross-section of charm-anticharm production in the $e^+e^-$-annihilation. It is well described by the following simple ansatz [1]:

$$\rho_{phen}(s) = s\left(f m_{res}^2 \delta(s - m_{res}^2) + \frac{1}{4\pi^2}\theta(s - s_o)\right) \qquad (3)$$

with the low-lying resonance of mass $m_{res} = m_{J/\psi}$, residue $f$ proportional to the electromagnetic width of the $J/\psi$-meson, and a smooth continuum spectrum with some effective threshold $s_o > m_{res}^2$. The experimental values of the spectrum parameters are:

$$4\pi^2 f \simeq 0.6\,, \quad m_{res} \simeq 3.1\, GeV\,, \quad s_o \simeq (4\, GeV)^2 \qquad (4)$$

---

[a] Talk at the workshop "Continuous Advances in QCD '96", Minneapolis, March 1996.

On the other hand this correlator is computable in perturbative QCD in the vicinity of vanishing momentum $q^2 = 0$ as that point is far away from the nearest threshold $4m_c^2$ due to charmed quark-antiquark pair. The parameter of the perturbative loop expansion here is $\alpha_s(4m_c^2) < 1$. Following [1] we consider the ratios $r_n = \mathcal{M}_{n+1}/\mathcal{M}_n$ of moments of the correlator (1). The moments are defined as

$$\mathcal{M}_n = \frac{1}{n!} \left\{ \left( -\frac{d}{dq^2} \right)^n \Pi_{\mu\mu}(q^2) \right\}_{q^2=0} \tag{5}$$

The applicability of perturbative QCD near $q^2 = 0$ implies the following expansion for the ratios $r_n$:

$$r_n = \frac{1}{4m_c^2} \left( a_n + b_n \alpha_s(4m_c^2) - c_n \frac{G^{(2)}}{(4m_c^2)^2} \right) \tag{6}$$

where $\{a_n, b_n, c_n\}$ are known numbers [1]. The term $\sim a_n$ comes from one loop of free charmed quarks. The term $\sim b_n$ comes from the two-loop diagrams corresponding to one-gluon exchange. The gluon condensate $G^{(2)} \equiv <0|(\alpha_s/\pi) G^a_{\mu\nu} G^a_{\mu\nu}|0>$ is the vacuum average of the first nontrivial operator in the Wilson operator-product expansion for the correlator of heavy-quark currents. The coefficient $c_n$ originating from the Wilson coefficient function starts with one loop of the heavy-quark propagators.

The lower ratios $r_{2,3,4}$, originating from short distances, are well reproduced in perturbative chromodynamics. The typical virtuality corresponding to the $n$th moment is $\sim 4m_c^2/n$. The coefficient $b_n$ grows with $n$. At large $n > 8$ the perturbation-theory based expansion (6) is irrelevant. The gluon condensate shows up in the intermediate ratios $r_{5,6,7}$ [1] sensitive to larger distances and, hence, nonperturbative fluctuations. The lower moments receive significant contribution from the high-energy (continuum) part of hadronic spectrum, whereas the higher moments correspond to long distances and are saturated by the lowest resonance $J/\psi$. The moment $\mathcal{M}_0$ is quadratically divergent and renormalizes the photon mass, whereas the moment $\mathcal{M}_1$ is logarithmically divergent and renormalizes the photon wave function. All the other moments are physical quantities: their ultraviolet singularities are absorbed by the renormalization of the QCD Lagrangian parameters, the charmed-quark mass and the strong coupling constant.

One can obtain a good estimate [1] of the renormalized QCD Lagrangian parameters by comparing the perturbative expressions for the lower ratios (6) to the same quantities obtained phenomenologically on the basis of dispersion relation (2). The principal uncertainty of this estimate is that the continuum contribution, significant in the lower ratios, is not known well enough from experiment. It is also only an approximation to identify the spectral density of the correlator of charmed-quark currents with the inclusive cross-section of the charm-anticharm production in $e^+e^-$-annihilation because of the contribution of light-quark currents to that inclusive cross-section. To avoid this problem and obtain an accurate way of calculating the renormalized charmed-quark mass and the strong coupling constant normalized on the charmed threshold it was suggested in [2,3] to calculate the correlator (1) by Monte-Carlo in Lattice QCD. Then the only experimental information needed is the low-lying resonance mass $(m_{J/\psi})$. It is used to fix the scale. This idea can only work if one can isolate physical contribution of the high-energy part of hadronic spectrum in Monte-Carlo data on the background of short-distance lattice artifacts. We demonstrate below that this is indeed the case.

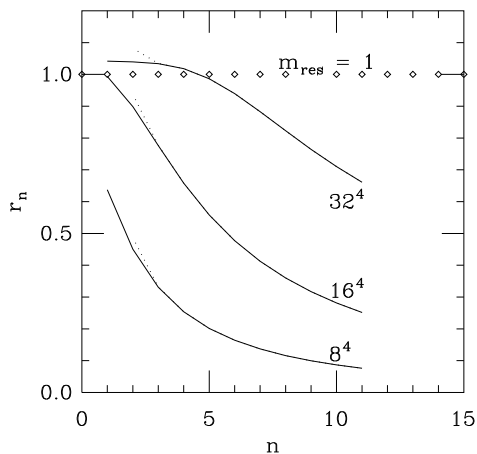

Figure 1: Ratios of the neighboring moments of the single resonance approximation to the correlator, on three lattices of different sizes: $8^4, 16^4, 32^4$. The resonance mass ($m_{res} = 0.5$) is given in the units of the inverse lattice spacing. Solid lines are to guide the eye. Dashed line shows the moments of the subtracted correlator. Horizontal dotted line is a prediction of the continuum theory.

## 2  Lattice artifacts of the moments

We plan to analyse Monte-Carlo data obtained on relatively small and coarse lattices. Hence the lattice artifacts is the number-one problem to deal with. As usual we distinguish finite-volume from the cut-off effects.

Rather than doing Fourier transform of the correlator (1) and evaluating derivatives of (5) on a discrete lattice we calculate the moments directly in the $X$-space:

$$\mathcal{M}_n = \frac{1}{2^{2n}\, n!(n+1)!} \int d^4x\, x^{2n}\, \Pi(x) \qquad (7)$$

We consider the correlator (1) in the infinitly-narrow-resonance approximation to be the propagator of a free field of mass $m_{res}$:

$$\Pi^{res}(q) = \frac{1}{m_{res}^2 + \frac{4}{a^2}\sum_\mu^4 \sin^2(q_\mu a/2)} \qquad (8)$$

The moments of a single resonance (8) computed with the help of eqn.(7) on lattices of different size are shown on Fig. 1. The straight horizontal line is expected in the continuum theory $r_n^{res} = 1/m_{res}^2$. One can see strong finite-volume effects on small lattices $8^4$, $16^4$ which persist even on the big (in a computational sense) $32^4$-lattice in the high moments. The higher moments originate in the long-distance part of the correlator. One can see the horizontal plateau of the continuum theory show up in the lower moments on the big $32^4$-lattice. On bigger lattices that plateau will expand to higher moments, but it will always remain somewhat above the prediction of the continuum theory due to cut-off $O(a^2)$-effects. Those cut-off effects in the moments of a single resonance are negligible for $a \cdot m_{res} < 1$ [3].

The lower moments of the correlator (1) are expected to be well reproduced in the loop expansion of perturbative QCD, where the dominant term is the one loop of free quarks with the renormalized mass $\bar{m}_c$. We need to know therefore the lattice artifacts of one loop of free Wilson fermions (as the Monte-Carlo data analysed was obtained for Wilson fermions). The moments of the correlator (1) in the one-loop (free quarks) approximation are shown in Fig. (2) for lattices of various sizes. Again one finds finite-volume effects which are particularly strong in high moments. Even on the big $32^4$-lattice the moments of one loop of Wilson quarks deviate from the continuum theory prediction (diamonds) due to the cut-off $O(a)$-effects. One can reconcile nicely the lattice moments (for $L \geq 32$) with the expectations of the continuum theory by using only a simple relation between the quark mass of the continuum theory $\bar{m}_c$ and the mass parameter of the Wilson propagator $m_c \equiv 1/2\kappa - 4$ [3]:

$$\bar{m}_c^2 = m_c^2 / (1 + m_c) \qquad (9)$$

All the masses in (9) are in the units of $a$. The mass $\bar{m}_c$ is defined from the behavior of the quark propagator near vanishing momentum. It has nothing to do with the pole mass.

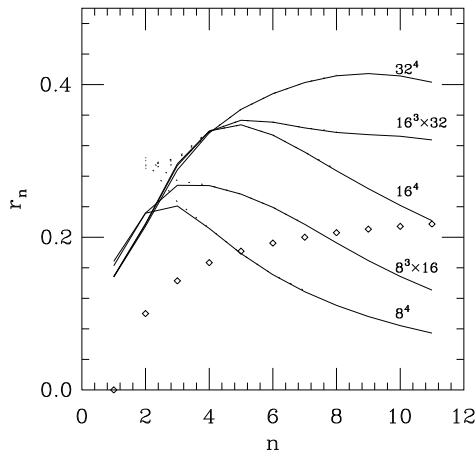

Figure 2: Ratios of the neighboring moments of the subtracted (dashed lines) and unsubtracted (solid lines) correlator of pseudoscalar currents, saturated by one loop of free Wilson fermions with $\kappa = 0.1$ ($am_c = 1$) and periodic boundary conditions, computed numerically on lattices of indicated sizes. Lines connect the data to guide the eye. The diamonds is a prediction of the continuum theory with the quark mass $\bar{m}_c \equiv m_c$.

Note that the relation (9) describing $O(a)$-fermionic cut-off effects is valid in the infinite volume. Corrections to this formula due to finite volume were studied in [3].

Strong finite-volume effects imply that one has to fit Monte-Carlo data not with the continuum-theory expressions for the moments but with the moments of a single resonance or free quarks, calculated in the same box where the Monte-Carlo simulation was done, which is what we do.

## 3  Dispersion relations on the Lattice

Our aim is to fit the short-distance part of the correlator of heavy-quark currents with the expressions of perturbative QCD. The short-distance part of a two-point correlator is determined to a great extent by the high-energy part of the spectrum. The problem therefore is to distinguish between the physical contribution of the high-energy part of spectrum and the pure lattice artifacts which will eventually dominate as one moves into the domain of very short distances. We claim to see a window of intermediate distances where the contribution of the high-energy part of sectrum is significant while the lattice artifacts do not dominate. Incorporating the high-energy part of the spectrum into the fit of Monte-Carlo data for the two-point correlators amounts to generalization of the dispersion relation to the lattice theory. The problem, in other words, is to construct the correlator of the lattice theory corresponding to a given spectral density.

We explore the following extention of the dispersion relation for the two-point correlator $\Pi(q^2)$ to the lattice theory:

$$\Pi(q^2) = \tilde{d}_1 + \tilde{d}_2 \frac{4}{a^2} \sum_\mu^4 \sin^2(q_\mu a/2) \tag{10}$$

$$+ \left(\frac{4}{a^2} \sum_\mu^4 \sin^2\left(\frac{q_\mu a}{2}\right)\right)^4 \int ds \frac{\rho(s)}{s^2 \left(s + \frac{4}{a^2} \sum_\mu^4 \sin^2(q_\mu a/2)\right)} \tag{11}$$

The factor $\sim sin^2$ is the inverse of the discrete Laplacian:

$$\int \frac{d^4q}{(2\pi)^4} e^{iqx} \frac{4}{a^2} \sum_\mu^4 \sin^2\left(\frac{q_\mu a}{2}\right) = \frac{1}{a^2} \sum_\mu^4 \left(2\delta(x) - \delta(x - \hat{a}_\mu) - \delta(x + \hat{a}_\mu)\right) \tag{12}$$

Therefore the coefficient $\tilde{d}_2$ in (10) is precisely the renormalization of the photon wave function in the case of vector currents of charmed quarks. The renormalized correlator would be given by eqn.(10), (11) with $\tilde{d}_1 = \tilde{d}_2 = 0$, considered as a function of the renormalized Lagrangian parameters (we also call it "subtracted correlator"). The subtraction constants $d_{1,2}$ are ultraviolet singular. In contrast to the continuum theory the term against $d_2$ in (10) is not a polynomial in $q^2$, so it contributes to all the moments, not only to $\mathcal{M}_1$. We find that its contribution to $\mathcal{M}_n$ decreases very fast with $n$. If the quark or resonance mass in units of $a$ is less than 1, the ratio $r_2$ is affected significantly by the term $\sim d_2$ (in contrast to the continuum theory, where this ratio is physical). The ratio $r_3$ is affected very little, whereas the contamination of all the other ratios $r_{n>3}$ is negligible. One can see the difference between the subtracted and unsubtracted correlators on Figs. (1) and (2). The short-distance lattice artifacts are confined to the ratios $r_{1,2}$ with rather small effect on $r_3$ as soon as the relevant masses in units of $a$ are less than 1. We therefore choose to fit the ratios $r_3$ and $r_4$ with the expressions of perturbative QCD.

## 4  Monte-Carlo data

The ratios of the moments of the correlator of vector currents of the continuum theory [1] are shown on Fig. 3. Higher moments of the correlator, obtained on the basis of the dispersion relation and experimental information about the spectral density, approach the moments of a single resonance ($J/\psi$-meson), which always lie on a straight horisontal line. The higher moments therefore come from long distances. The lower moments ($r_{2,3}$) are seen to be fitted well by the moments of one loop of free charmed quarks, they are short-distance quantities. The two-loop correction due to one-gluon exchange will improve agreement between perturbative predictions and experimental predictions of the lower moments of the correlator. Incorporation of the gluon condensate will improve fit in somewhat higher moments $r_{5,6,7}$ [1].

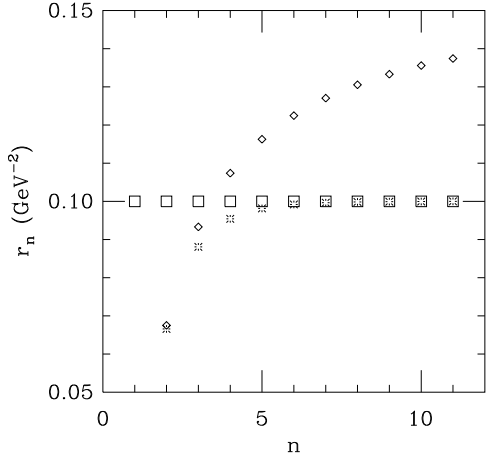

Figure 3: Ratios of the neighboring moments in the continuum theory for the correlator of interpolating currents of the $J/\psi$-meson. Burst symbols correspond to the phenomenological evaluation of the correlator via dispersion relation with the smooth hadronic continuum spectrum modelled as in [2]. Squares show the contribution of a single $J/\psi$-resonance. Diamonds are moments of one loop of free quarks of mass $m_c = 1.26\ GeV$.

Compare Fig. 3 with Fig. 4, which shows the ratios of the successive moments of the same correlator computed by Monte-Carlo on $16^3 \star 32$-lattice for $\beta = 6$, $\kappa = 0.1100$ with the tadpole-improved Wilson Clover action [4]. One can see strong finite-volume effects in Monte-Carlo data in the domain of higher moments. The moments of a single-resonance (solid line) exhibit the same finite-volume effects and fit Monte-Carlo very well in a wide range of higher moments. This allows us to fix the scale of the lattice theory.

The lower moments $\{r_n, n < 6\}$ are seen to deviate strongly from the single-resonance curve. All Monte-Carlo moments are fitted remarkably well by the dashed line. This line shows the moments of the lattice correlator corresponding to the spectral density which incorporates a smooth continuum spectrum in addition to the low-lying resonance. That lattice correlator is obtained using the disperion relation (11). The parameters of the spectral density are chosen to have experimental values as in (4). Although the lattice artifacts in Monte-Carlo data are strong, they are well described by the natural modification of the denominator in (11), which describes propagation of a state of a given invariant mass $s$ on a finite lattice. We find evidence that it is a good approximation to ignore possible lattice artifacts in the spectral density, as it is done in (11). Incorporation of the high-energy (continuum) part of hadronic spectrum changes the line of ratios $r_n$ dramatically in small $n$. As expected the high-energy part of spectrum is important to descibe the short-distance behaviour of the correlator. It is less trivial, that one can identify the physical contibution of the continuum spectrum and distinguish it from the short-distance lattice artifacts, which are confined to the lowest ratios $r_{1,2}$, on the lattices of reasonable size.

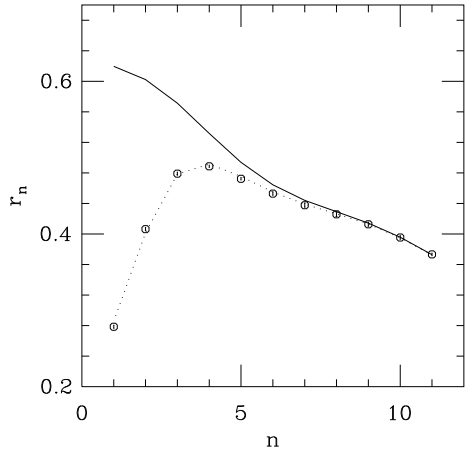

Figure 4: Monte-Carlo data on the $16^3 \times 32$, $\beta = 6$ lattice for vector currents. $\kappa = 0.1100$. The solid line is a single resonance approximation. The dashed line has been obtained with the phenomenological charmonium spectrum incorporated via the dispersion relation.

## 5   How to fit Monte-Carlo data at short times

The successfull incorporation of the continuum spectrum helps to fit Monte-Carlo data on shorter distances and hence fix the scale on small lattices. Fig.

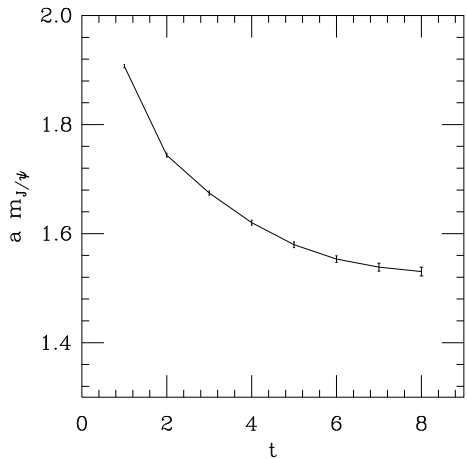

Figure 5: Resonance mass from the conventional fit of the zero-momentum component of the correlator of vector currents to $cosh[(t - L/2)m_{res}]$ on the smaller $8^3 \times 16$ lattice.

5 shows the conventional way of extracting the lowest-resonance mass from the two-point correlator of local currents on the small $8^3 \star 16$-lattice: Monte-Carlo data are fitted by the expression for the correlator in the signle-resonance approximation. Horizontal plateau is expected at large times which height is given by the resonance mass. That plateau is not seen on Fig. 5 because the $8^3 \star 16$ lattice is too small. The same type of picture holds (see Fig. 6) in terms of the moments: the resonance mass exptracted from the *single-resonance fit* to different moments should exhibit a horizontal plateau. On the $8^3 \star 16$ lattice that plateau only starts to form at higher moments. At lower moments the resonance mass deviates from the horizontal line. This deviation is due to significant contribution of the high-energy part of hadronic spectrum at short distances.

We now include the continuum spectrum as in eqn. (4) via the dispersion relation (11) keeping the ratio of the effective continuum threshold to the resonance mass fixed to have the experimental value $s_o/m_{res}^2 \approx 1.7$ and use the resonance mass as the only fit parameter. The result of such a fit is shown on Fig. 6. One can see that the incorporation of the continuum spectrum via the dispersion relation (11) allows one to fix the resonance mass unambiguously from the data on relatively short distances.

Proceed now to a bigger $16^3 \star 32$ lattice, where one can determine the resonance mass by the conventional fit anumbiguously since the time extent is large enough for the high-energy excitations to decay. The ground state dominates and constitutes horizontal plateau at $t \geq 10$. The resulting resonance masses are shown in Tables 2, 3. One can compare them to the resonance masses extracted from the small lattice by means of the dispersion relation (incorporating the continuum spectrum), shown in Table 1. The agreement is very good. The corresponding spectrum of charmonium bound states is shown on Fig. 8.

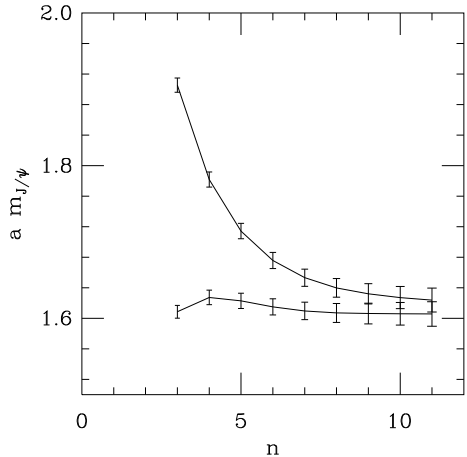

Figure 6: Effective resonance mass from fitting the moment ratios of Monte-Carlo data on the $8^3 \times 16$, $\beta = 6$ lattice for vector currents. $\kappa = 0.1060$. The upper line is a single-resonance approximation.

## 6 The renormalized charmed-quark mass

We fit the lower moment ratios $r_{3,4}$ of Monte-Carlo data with the free-quarks approximation to the correlator (1):

$$r_3(\text{free Wilson quarks}) = r_3(\text{Monte-Carlo}) \quad (13)$$

The Wilson parameter $\kappa$ corresponding to those free quarks determines the the renormalized quark mass in accordance with (9) if Monte-Carlo is done on sufficiently large lattice, such as $16^3 \star 32$. On smaller lattices the relation (9) needs to be corrected due to finite-volume effects. This is done by means of the dispersion relation [3].

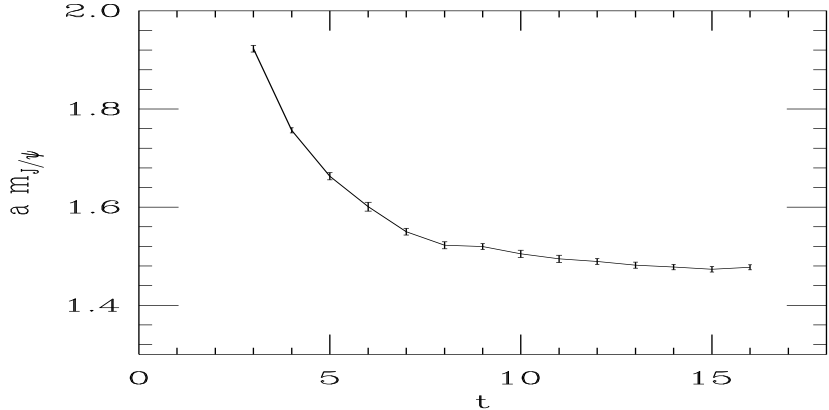

Figure 7: Resonance mass from the conventional fit of the zero-momentum component of the correlator of vector currents to $cosh[(t-L/2)m_{res}]$ on the bigger $16^3 \times 32$ lattice.

In order to fix the subtraction scheme one should be sensitive to the $\alpha_s$-correction in the ratios $r_n$. The study of the lattice artifacts on the two-loop level is not done yet. However, since the term $b_n\,\alpha_s$ of eqn. (6) is small in the lower ratios $r_{3,4}$, its value may be taken from the continuum theory:

$$\begin{aligned} r_n &= \frac{a_n}{4m_c^2}\,\xi_n \\ \xi_n &\equiv 1 + \alpha_s\,b_n/a_n \end{aligned} \qquad (14)$$

We take the value of the strong coupling constant $\alpha_s(m_c) \approx 0.3$ and the coefficients $b_n/a_n$ - from the two-loop calculations of the continuum theory[1] for the vector, pseudoscalar and scalar channels:

$$\begin{aligned} \xi_3^{psc} &= 1.02\,, \quad \xi_3^{vec} = 0.95\,, \quad \xi_3^{sca} = 0.97 \\ \xi_4^{psc} &= 0.96\,, \quad \xi_4^{vec} = 0.92\,, \quad \xi_4^{sca} = 0.91 \end{aligned} \qquad (15)$$

The values of $\xi_n$ in (15) correspond to our choice of the coefficients $b_n$ in the $\bar{M}S$-subtraction scheme with the normalization point $\mu = m_c$ [5]. After the fit of Monte-Carlo data with free fermions (one-loop approximation) the mass of those fermions must corrected by the coefficients $\xi$ in accordance with (15).

The results for the renormalized charmed-quark mass are shown in Tables 1, 2 , 3. The mass is extracted from the ratios $r_3$ and $r_4$ independently in three different channels: pseudoscalar, vector and scalar; on two lattices of size $8^3 * 16$ and $16^3 * 32$, for a few values of the hopping parameter $\kappa$ in the vicinity of $\kappa_{charm}$ of which $\kappa = 0.1060$ (for $\beta = 6$) and $\kappa = 0.1150$ (for $\beta = 6.3$) are shown. The values of $\kappa$ correspond to the tadpole-improved Wilson Clover action [4]. The data at $\beta = 6.3$ checks the scaling properties of the renormalized heavy-quark mass. By definition this physical mass should stay fixed and finite as $a \to 0$ ($\beta \to \infty$).

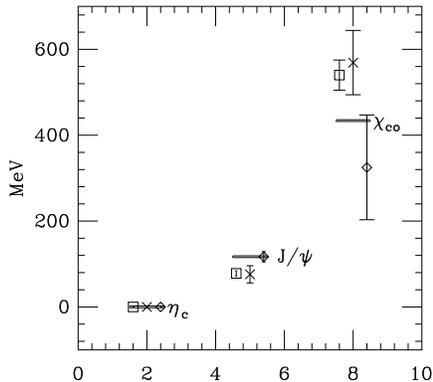

Figure 8: Charmonium spectrum obtained on the $16^3 \times 32$ lattice at $\beta = 6$ (squares) and $\beta = 6.3$ (diamonds) and on the small lattice $8^3 \times 16$ at $\beta = 6$ (crosses) with the clover-and-tapole-improved action. The pseudoscalar mass $m_{\eta_c} = 2.979 GeV$ is used for normalization. The wide horizontal lines are experimental data.

One can see a rather stable value of the renormalized charmed-quark mass $m_c^{\overline{MS}}(m_c) = 1.22(5) GeV$, which is well in agreement with the estimates of the continuum theory [6]: $m_c^{\overline{MS}}(m_c) \approx 1.23 GeV$. The previously reported lattice result is [7]: $m_c^{\overline{MS}}(m_c) = 1.5(3) GeV$.

## 7 Conclusions

Study of the short-distance behavior of the correlator of heavy-quark currents on the lattice appeares as a very promising way to accurately calculate the parameters of perturbative QCD.

### Acknowledgments

This report is based on work done in collaboration with Ph. de Forcrand. Computer time for this project was provided by the Minnesota Supercomputer Institute and by the Pittsburgh Supercomputer Center.

Table 1: The charm-quark mass obtained from fitting the ratios $r_3$, $r_4$ of the subtracted correlator on the small lattice $8^3 \times 16$ for $\beta = 6$, $\kappa = 0.1060$. The masses $m_{res}$ and $\bar{m}_c$ are in units of the lattice spacing. The masses $m_c^{\overline{MS}}[a_{J/\psi}]$ and $m_c^{\overline{MS}}[a_{\eta_c}]$ are in $GeV$. They are obtained assuming that the lattice spacing is fixed from the vector channel and from the pseudoscalar channel respectively ($a^{-1} \approx 1.9 GeV$). The indicated errors are statistical.

| fit to $r_3$ | $m_{res}$ | $\bar{m}_c$ | $\bar{m}_c/m_{res}$ | $m_c^{\overline{MS}}[a_{J/\psi}]$ | $m_c^{\overline{MS}}[a_{\eta_c}]$ |
|---|---|---|---|---|---|
| $\eta_c$ | 1.57(1) | .596(2) | .381(2) | 1.16(1) | 1.15(1) |
| $J/\psi$ | 1.61(2) | .651(4) | .405(3) | 1.22(1) | 1.21(1) |
| $\chi_o$ | 1.87(10) | .609(9) | .329(10) | 1.15(3) | 1.14(3) |
| fit to $r_4$ | $m_{res}$ | $\bar{m}_c$ | $\bar{m}_c/m_{res}$ | $m_c^{\overline{MS}}[a_{J/\psi}]$ | $m_c^{\overline{MS}}[a_{\eta_c}]$ |
| $\eta_c$ | 1.57(1) | .623(3) | .398(2) | 1.17(1) | 1.16(1) |
| $J/\psi$ | 1.61(2) | .666(4) | .415(3) | 1.23(1) | 1.22(1) |
| $\chi_o$ | 1.87(10) | .649(10) | .350(10) | 1.19(3) | 1.18(3) |

Table 2: Same as in Table 1 for the big lattice $16^3 \star 32$, $\beta = 6$, $\kappa = 0.1060$.

| fit to $r_3$ | $m_{res}$ | $\bar{m}_c$ | $\bar{m}_c/m_{res}$ | $m_c^{\overline{MS}}[a_{J/\psi}]$ | $m_c^{\overline{MS}}[a_{\eta_c}]$ |
|---|---|---|---|---|---|
| $\eta_c$ | 1.562(5) | .618(2) | .396(2) | 1.208(6) | 1.190(6) |
| $J/\psi$ | 1.600(6) | .660(1) | .413(2) | 1.246(6) | 1.228(6) |
| $\chi_o$ | 1.86(4) | .645(2) | .347(2) | 1.230(8) | 1.212(8) |
| fit to $r_4$ | $m_{res}$ | $\bar{m}_c$ | $\bar{m}_c/m_{res}$ | $m_c^{\overline{MS}}[a_{J/\psi}]$ | $m_c^{\overline{MS}}[a_{\eta_c}]$ |
| $\eta_c$ | 1.562(5) | .645(1) | .413(2) | 1.224(6) | 1.206(6) |
| $J/\psi$ | 1.600(6) | .677(1) | .424(2) | 1.258(6) | 1.240(6) |
| $\chi_o$ | 1.86(4) | .686(3) | .369(7) | 1.27(1) | 1.25(1) |

Table 3: Same as in Table 2 at $\beta = 6.3$, $\kappa = 0.1150$. The lattice spacing : $a^{-1} \approx 3.7 GeV$

| fit to $r_3$ | $m_{res}$ | $\bar{m}_c$ | $\bar{m}_c/m_{res}$ | $m_c^{MS}[a_{J/\psi}]$ | $m_c^{MS}[a_{\eta_c}]$ |
|---|---|---|---|---|---|
| $\eta_c$ | 0.805(6) | .307(2) | .381(2) | 1.149(6) | 1.148(6) |
| $J/\psi$ | 0.836(7) | .336(2) | .401(3) | 1.214(9) | 1.212(9) |
| $\chi_o$ | 0.89(6) | .338(5) | .38(2) | 1.23(3) | 1.23(3) |
| fit to $r_4$ | $m_{res}$ | $\bar{m}_c$ | $\bar{m}_c/m_{res}$ | $m_c^{MS}[a_{J/\psi}]$ | $m_c^{MS}[a_{\eta_c}]$ |
| $\eta_c$ | 0.805(6) | .321(2) | .399(2) | 1.165(6) | 1.164(6) |
| $J/\psi$ | 0.836(7) | .344(2) | .411(2) | 1.222(6) | 1.222(6) |
| $\chi_o$ | 0.89(6) | .359(8) | .40(2) | 1.27(4) | 1.27(4) |